\documentstyle[aps,12pt]{revtex}

\newcommand\lsim{\mathrel{\rlap{\lower4pt\hbox{\hskip1pt$\sim$}}
	\raise1pt\hbox{$<$}}}
\newcommand\gsim{\mathrel{\rlap{\lower4pt\hbox{\hskip1pt$\sim$}}
	\raise1pt\hbox{$>$}}}

\newcommand\be{\begin{equation}}
\newcommand\ee{\end{equation}}
\newcommand\eq{\begin{equation}}
\newcommand\en{\end{equation}}
\newcommand\bea{\begin{eqnarray}}
\newcommand\eea{\end{eqnarray}}
\newcommand\nn{\nonumber}
\newcommand\lt{\left}
\newcommand\rt{\right}

\newcommand\GeV{{\, \mathrm{GeV}}}
\newcommand\mpl{M_{\rm Pl}}

\newcommand\Vsb{V_{\rm susy \!\!\!\!\!\!\!\!\!\!
        \raisebox{-2.25pt}{\line(3,2){12}} \,\,}}

\begin{document}

\title{Inflationary models with a flat potential enforced by \\
 non-abelian discrete gauge symmetries}
\bigskip

\author{ E. D. Stewart\footnote{ewan@kaist.ac.kr}}
\address{NASA/Fermilab Astrophysics Group, FNAL, Batavia, IL 60510 \\
and Department of Physics,
KAIST, Taejon 305-701, Korea\footnote{Address from August 1999}}

\author{J. D. Cohn\footnote{jcohn@cfa.harvard.edu}}
\address{Harvard Smithsonian Center for
Astrophysics,
60 Garden St., Cambridge, MA, 02138}

\date{\today}
\maketitle
\begin{abstract}
Non-abelian discrete gauge symmetries can provide the inflaton with a flat
potential even when one takes into account gravitational strength effects.
The discreteness of the symmetries also provide special field values where
inflation can end via a hybrid type mechanism.
An interesting feature of this method is that it can naturally lead to
extremely flat potentials and so, in principle, to inflation at unusually
low energy scales.
Two examples of effective field theories with this mechanism are given,
one with a hybrid exit and one with a mutated hybrid exit.
They include an explicit example in which the single field consistency
condition is violated.
\end{abstract}
%\pacs{98.80.Cq}

\vspace{-100ex}
{\raggedleft FERMILAB-Pub-99/188-A \\}
{\raggedleft KAIST-TH 99/9 \\}
\vspace{-\baselineskip}
\vspace{100ex}

\thispagestyle{empty}
\setcounter{page}{0}
\newpage
\setcounter{page}{1}

\section{Introduction}

Inflation \cite{Gliner} explains many basic features of our universe
\cite{Guth,early}.
It is also thought to have generated the density perturbations needed to
form galaxies and all the other large scale structure in the observable
universe \cite{perturbations}.
There are many types of inflation that are natural from the particle
physics point of view.

If the energy density of our vacuum (the cosmological constant) is
positive,
then it will eventually give rise to inflation.
Observations suggest this is just beginning now.
Although this type of inflation is natural, the fact that it is just
beginning
{\em now\/} can (in the opinion of EDS) only be explained by anthropically
selected fine-tuning of the cosmological constant.

If in the past the universe became trapped in a positive energy false
vacuum for sufficiently long, one will get an epoch of false vacuum
(old) inflation \cite{Guth}.
This probably did happen, though in the unobservably distant past.
A false vacuum with near Planck scale energy density could start
(eternal) inflation from fairly generic initial conditions.
The desirable properties (and maybe even necessity) of eternal inflation
have been stressed by Linde \cite{linde} in the context of $\phi^n$
chaotic inflationary potentials.
Unfortunately, such potentials generically do not survive the inclusion
of gravitational strength effects,
especially for the extremely large field values needed to start eternal
inflation at the Planck density.
However, much the same ideas can be realized using the generic false
vacuum inflation.

Thermal inflation \cite{thermal} just needs a potential
$V = V_0 - \frac{1}{2} m^2 \phi^2 + \ldots$ with $m \ll V_0^{1/4}$,
typical of supersymmetric theories.
It occurs when $\phi$ is held at $\phi=0$ by thermal effects, and is
probably
needed to solve \cite{thermal} the moduli (Polonyi) problem
\cite{polonyi}.
It also has important implications for baryogenesis and dark matter
\cite{tib,tia,tib2,tim,tis}.

Rolling scalar field inflation just needs a potential
$V = V_0 - \frac{1}{2} m^2 \phi^2 + \ldots$ with $m \sim V_0^{1/2} / \mpl$
where $\mpl = 1/\sqrt{8\pi G} \simeq 2.4 \times 10^{18} \GeV$, typical of
moduli potentials.
It occurs as the inflaton $\phi$ rolls off the maximum of the potential.
This may also have happened.

However,  observations constrain the density perturbations to be
approximately scale-invariant.
Therefore, the natural way to produce these is with an approximately
scale-invariant inflation.
The only known scale-invariant inflation is a limit of rolling scalar
field
inflation called slow-roll inflation \cite{slowroll}.
It requires the stronger condition $m \ll V_0^{1/2} / \mpl$, or more
generally
\be
\label{vpcond}
\left( \frac{V'}{V} \right)^2 \ll \frac{1}{\mpl^2}
\ee
and
\be
\label{vppcond}
\left| \frac{V''}{V} \right| \ll \frac{1}{\mpl^2}
\ee
The first condition suggests we should be near a maximum, or other
extremum, of the potential.
The second is non-trivial \cite{fvi,iss}.
For example, many models of inflation are built ignoring gravitational
strength interactions, and so are implicitly setting
$ M_{\rm Pl} = \infty $.
Clearly one cannot achieve the second condition in this context.
In supergravity, the potential is composed of two parts, the $F$-term and
the $D$-term.
If the inflationary potential energy is dominated by the $F$-term then one
can show that \cite{Dine,fvi,iss}
\be
\label{etaprob}
\frac{V''}{V} = \frac{1}{\mpl^2} + \mbox{model dependent terms}
\ee
Unless the model dependent terms cancel the first term, the second slow
roll condition, Eq.~(\ref{vppcond}) above, is violated.
Thus to build a model of slow-roll inflation one must be able to control
the gravitational strength corrections.

There have been various attempts at achieving slow-roll inflation
naturally, which are summarized below.
For extensive references on inflationary models, see, for example,
\cite{ly-rio}.

Special forms for the Kahler potential \cite{fvi,iss,Gaillard}:
The $F$-term part of the potential is determined by the
superpotential $W$ and the Kahler potential $K$.
The Kahler potential contains most of the terms which
make slow-roll inflation difficult.
Choosing a special form for the Kahler potential combined with some
other conditions can allow one to cancel off the model independent
gravitational strength corrections that generically destroy slow-roll
inflation.
Kahler potentials of the required form arise in large radius,
weak coupling limits of string theory or in models with some effective
extended supersymmetry.

D-term domination of the inflationary potential energy
\cite{dterm}\footnote{The first $D$-term model of inflation was given
in \cite{Casas} but the model and the motivation were different.}:
Naively simple, but in order to obtain the COBE normalisation one must
stabilize a modulus at a very large value without the aid of $F$-term
supersymmetry breaking.

Flattening the inflaton's potential with quantum corrections
\cite{flattening,Covi}:
This is completely natural but is being tested by observations
and may not succeed.

Cancellation mechanism \cite{Ross}: Here the expectation value of a
Nambu-Goldstone boson is used to cancel the inflaton's mass to produce
slow-roll.

In this paper we use non-abelian discrete gauge symmetries to
guarantee the flatness of the inflaton's potential.
The basic idea was presented in \cite{letter}.  Here two
full inflationary models utilizing this idea are constructed,
a hybrid model and a mutated hybrid model.
The inflationary mechanism requires the inclusion of higher order
terms in the superpotential (and Kahler potential
and supersymmetric loop corrections), and quantitative calculation of
the properties of the exit.
As the hybrid model can have a very flat potential, it can have a low energy
scale, but this also brings with it the possibility of large fluctuations
\cite{spike} during the exit which provides a stringent constraint.
This inflationary mechanism has the advantage that one can
work in the low energy effective
field theory, without needing to know the detailed high energy theory.

In Section~\ref{review} we briefly review the construction of a low energy
effective supergravity theory; see textbooks, for example
Ref.~\cite{bailin}, and references therein for further information.
Readers familiar with low energy effective supergravity model building
can skip this section.
In Section~\ref{idea} we describe our basic idea.
In Sections~\ref{hybrid} and~\ref{mutated} we give examples of models
implementing this idea.
In Section~\ref{con} we give our conclusions.
In the Appendix we list useful properties of the non-abelian discrete
group $\Delta$(96) that we use to build the models of
Sections~\ref{hybrid} and~\ref{mutated}.

\section{Review of low energy effective supergravity model building}
\label{review}

The scalar potential of a supergravity theory is specified by
its full Kahler potential $K(\phi_i,\bar{\phi}_i)$,
superpotential $W(\phi_i)$ and $D$-terms.
We use discrete gauge symmetries which have
no associated gauge fields and hence no $D$-terms.
In the full supergravity theory, the scalar potential is 
\eq
V(\phi_i) = e^K \left[ \left( W_{\phi_i} + W K_{\phi_i} \right)
K^{-1}_{\phi_i \bar{\phi}_j}
\left( \bar{W}_{\bar{\phi}_j} + \bar{W} K_{\bar{\phi}_j} \right)
- 3 |W|^2 \right] + \mbox{$D$-terms}
\en
where the $\{ \phi_i \}$ include fields in all sectors,
hidden and not hidden.
As there are no $D$-terms in our case we will not
discuss them further.

Writing $\phi$ for the set of fields $\{ \phi_i \}$,
only the combination
\eq
G(\phi,\bar{\phi}) = K + \ln |W|^2
\en
is physically relevant, and so the freedom to make a 
Kahler transformation remains
\eq
K(\phi,\bar{\phi}) \to K(\phi,\bar{\phi}) - F(\phi) - \bar{F}(\bar{\phi})
\en
\eq
W(\phi) \to e^{F(\phi)} W(\phi)
\en
Thus the Kahler potential can be chosen to be independent of
holomorphic and anti-holomorphic terms.
The kinetic term is
\eq
K_{\phi_i \bar{\phi}_j} \partial_\mu \phi_i \partial^\mu \bar{\phi}_j
\en
and so, for a standard kinetic term, the leading term in the
Kahler potential will be
\eq
K = \bar{\phi}_i \phi_i + \mbox{higher order terms}
\en
The superpotential $W$ consists of all holomorphic terms allowed
by the symmetries.
This is the expression in terms of all the fields in the theory.

For the effective field theory,
only some of the fields are dynamical, and the rest are integrated
out.  The symmetries will dictate allowed terms for the dynamical
fields.
To leading order
(in the flat space, i.e. the $M_{\mathrm{Pl}} \to \infty$ limit), 
the potential for the dynamic fields (written here as $\phi_i$ as well)
in the effective field theory is given by
(see Eq.~7.5 of Ref.~ \cite{nilles})
\begin{eqnarray}
V_{\mathrm{low\ energy,\ leading}}(\phi_i)
&=&
V_0 +\sum_j \left| \frac{\partial W}{\partial \phi_j} \right|^2 +
m_j^2 \left|\phi_j\right|^2 + \mu \tilde{W} + \mbox{c.c.} \nn \\
&=&
\sum_j \left| \frac{\partial W}{\partial \phi_j} \right|^2 + \Vsb
\end{eqnarray}
(Ref.~\cite{nilles} discusses the presence of a possible cosmological
constant which in our case is the vacuum energy during inflation and
is taken nonzero in order for inflation to occur.)
The combination $\tilde{W}$ is an expansion in the dynamical fields
obeying the symmetries and holomorphicity just as $W$ does, but the
coefficients of the various allowed terms are different from those in $W$
(the low energy coefficients are induced by integrating out terms with the
heavier fields and thus do not have any fixed relation in the low energy
theory).
The coefficient $\mu$ comes from $\langle W_{\phi_{\mathrm{hid}}} \rangle$
and so is naively the size of
$|F_{\mathrm{hid}}| = |W_{\phi_{\mathrm{hid}}} + K_{\phi_{\mathrm{hid}}} W|$.
The masses $m_j^2$ can be positive or negative with a magnitude
$\lt| m_j^2 \rt| \gsim |F_{\mathrm{hid}}|^2 \gsim V_0$. 
See \cite{nilles,bailin} and references therein for more discussion.

Before going on to include higher order terms in $M_{\mathrm{Pl}}$,
it is useful to compare the scale of supersymmetry breaking
and the potential energy scale.
Having supersymmetry breaking in the full theory means
that  $|F_{\phi_i}| =  |W_{\phi_i} + W K_{\phi_i}| \ne 0$
for some $\phi_i$.  We will for simplicity take this to be for
one field $\phi_i$ and so write $|F|$ for the scale of supersymmetry
breaking.  This field is in the hidden sector by
construction.
One has in the full theory during inflation
\eq
V = V_0 > 0 \leftrightarrow |F|^2 \sim 
|W_{\phi_{hid}} + W K_{\phi_{hid}}|^2 > 3 |W|^2
\en
(the Kahler potential merely multiplies this out front).
With a vacuum energy, $V= V_0 > 0$, as is the case during 
inflation, one has in the full formula that 
$|F| \gsim |W|$ (for a positive potential) as well as
$|F|^2 \gsim V_0$ (as it is decreased by $3|W|^2$).
Note that $V_0$ does not have to be the dominant source of
susy breaking in our models (\mbox{i.e.} inflation can occur below the
scale of susy breaking), so $V_0$ does not necessarily characterize
the scale of the susy breaking parameters. This
large degree of freedom and uncertainty makes it difficult to
relate the scale of supersymmetry breaking and the vacuum energy
in a model independent way.

To include the higher order possible terms in the theory, we have
contributions from the visible and hidden sector.
The leading terms from supergravity are determined by leading
terms in the expansion of the superpotential $W$ and
$\Vsb$, both of which will be given for each model.
The higher order terms nonzero around the background will
be discussed as they appear with the exception of 
the leading term in
$V_0 K$, which has already been included in the
masses.  These higher order terms include terms of the form
$|W|^2$,
$W_\phi K_{\bar{\phi}} \bar{W} + c.c.$
and $| K_{{\phi}} {W}|^2$.
By construction $F_{\mathrm{hid}}$ has a 
constant leading term (for supersymmetry breaking) plus, in terms of the low
energy effective theory fields, some expansion in the low
energy fields.  The other terms will also be expansions in
the low energy fields, and all of these must obey the symmetries
(as the symmetries do not mix the hidden and visible sectors by
construction).
The leading terms are thus
\begin{eqnarray}
V &=& \sum_i \lt| W_{\phi_i} \rt|^2
+ |F|^2 \times \mbox{(arbitrary real function of the fields)} \nn \\
&& \mbox{} + \langle W \rangle \times \mbox{(antiholomorphic function)}
+ \mbox{c.c.}
\end{eqnarray}
which can each be multiplied by some arbitrary real function
(e.g. from the expansion of the Kahler potential).
The higher order terms are usually irrelevant because they are only small
corrections to existing terms.

It should be kept in mind that this effective field theory is likely
to correspond to a
different region of field space than we are in now.  In particular,
although there are expected values for vacuum energy and supersymmetry
breaking today, these numbers correspond to expansion of the
full high energy theory around the background we have today.  In this
earlier time, the vacuum energy and supersymmetry breaking terms will
correspond to an expansion around a different point in field space, and
thus may have very different values.
As the slow roll conditions are violated, inflation ends, and 
the field values begin to change by large amounts. 
This takes the theory outside the realm of the validity of
the effective field theory under consideration, which is an expansion around
a particular point in field space.
Thus, although there will be contributions to the potential
that will be seen to subtract from the vacuum energy $V_0$,
the ultimate value of the cosmological constant after this inflationary
era is not determined within this theory.  That is, although the
cosmological constant is decreasing when inflation ends, the subsequent
behavior of the fields causing this cancellation (and thus some specified
value of $V_0$ after inflation ends) requires knowledge of the potential
for these fields outside of the regime of the effective field theory.
The very important question of the observed cosmological constant today
is a major outstanding problem in theoretical particle physics and
outside the issues of concern in this paper.
This value is a combination
of the value reached after the inflationary stage described here (determined
by the minimum of the potential of the full theory at the end of
inflation, unspecified in
an effective field theory) plus any other contributions which arise as the 
universe evolves to its current day temperature and field configuration.
More (although less comprehensive) 
information is needed about the full theory as well in order to specify 
what happens next, in particular details of (p)reheating (which involves
fields which were not dynamical during inflation and thus did
not appear in the effective field theory), or which
fields are dynamical in the next era, etc.  

\section{The idea}\label{idea}

One of the better early attempts to naturally achieve a flat inflaton
potential was Natural Inflation \cite{natural}.
The inflaton was the pseudo-Nambu-Goldstone boson $\theta$ of an
approximate U(1) global symmetry.
The potential was of the form
\be
V = \epsilon f\!(\theta)
\ee
where $\epsilon \rightarrow 0$ in the limit of exact symmetry.
Thus the inflaton's mass
\be
V'' \propto \epsilon
\ee
can be made arbitrarily small.
However, in this model one can {\em not\/} use the U(1) global symmetry
to enforce
\be
\lt| \frac{V''}{V} \rt| \ll \frac{1}{\mpl^2}
\ee
because $V$ also vanishes in the limit where the symmetry is exact.

This problem can be solved by adding a constant to the potential
\be
V = V_0 + \epsilon f\!(\theta)
\ee
in which case one could in principle make $|V''/V|$ arbitrarily small.

However, one must now find a way to end inflation.
Inflation can end if there is some critical value of the inflaton,
$\theta = \theta_{\rm c}$, at which the potential destabilizes.
For example, this could happen due to a term in the potential of the form
$[\lambda^2\phi_0^2\sin^2(n\theta)-m^2]\psi^2$ with $\lambda\phi_0 > m$.
For $\theta < \theta_{\mathrm{c}} = (1/n)\sin^{-1}(m/\lambda\phi_0)$,
the potential is unstable to $\psi \rightarrow \infty$
and the runaway of $\psi$ can cancel out the vacuum energy $V_0$.
See Refs.~\cite{hybrid,fvi} for general discussion of the hybrid inflation
mechanism.
This critical value must violate the U(1) symmetry,
as a particular value of $\theta$ is singled out.
However, special values of $\theta$ can be consistent with a
discrete subgroup of the U(1) symmetry being unbroken,
${\bf Z}_{2n}$ in the above example.

Furthermore, if this discrete subgroup is gauged, it can be regarded as
fundamental, with the approximate U(1) global symmetry arising as a
consequence.
For example, if one had fields $\phi_+$ and $\phi_-$ with charges
$+1$ and $-1$ respectively under a ${\bf Z}_4$ symmetry,
then the lowest dimension (and thus dominant) invariants,
$\phi_+ \phi_-$, $\lt| \phi_+ \rt|^2$, and $\lt| \phi_- \rt|^2$,
are invariant under the extended global U(1) symmetry, while terms which
explicitly break the U(1), such as $\phi_+^4$, are of higher order.
The exact discrete ${\bf Z}_4$ symmetry thus gives rise to an approximate
U(1) symmetry in the region of field space in which $\lt|\phi_+\rt|$ and
$\lt|\phi_-\rt|$ are small.

In order to realize the couplings necessary for the hybrid inflation
mechanism, for example $\lambda \phi^2 \psi^2$, it is more natural to use
a non-abelian discrete symmetry.
The inflaton then corresponds to the pseudo-Nambu-Goldstone bosons,
$\Phi_a / |\Phi|$, of the approximate non-abelian continuous symmetry,
and the hybrid exit is implemented when the {\em magnitude\/} of one of
the components of a representation of the symmetry reaches some critical
value, for example $\lt| \Phi_1 \rt| = \Phi_{\rm c}$, rather than when the
phase of a field reaches some critical value, which would be the case if
one were to use an abelian discrete symmetry.

For a (discrete) gauge theory to be consistent it must be anomaly free
\cite{Ibanez}.
However, only the linear anomaly conditions survive for discrete abelian
gauge symmetries \cite{Banks}.
For the same reasons we expect only linear anomaly conditions to survive
for non-abelian discrete gauge symmetries.
However, there are no linear anomaly conditions for non-abelian gauge
symmetries.
Therefore, by this argument, non-abelian discrete gauge symmetries should
be automatically anomaly free.
Of course, any other gauge symmetries in the model will have to satisfy
the usual anomaly conditions.

In order to have our pseudo-Nambu-Goldstone bosons, we need a potential
which {\em spontaneously\/} breaks the extended continuous symmetry,
fixing $|\Phi| \equiv \lt(\sum_a \lt|\Phi_a \rt|^2 \rt)^{1/2}$ at some
value $\Phi_0 > 0$.
Our non-abelian discrete gauge symmetry is then non-linearly realized
on the pseudo-Nambu-Goldstone bosons which include the inflaton.
This protects the inflaton mass from large corrections.
In this paper, we assume a hidden sector breaks supersymmetry.
This generates supersymmetry breaking terms, including a vacuum energy
$V_0$ and masses for the scalars, in our effective potential.
We then use the renormalization group running of the supersymmetry
breaking mass term for $\Phi$ to generate a potential for $\Phi$ with
non-trivial minimum $|\Phi| = \Phi_0$ \cite{bailin}.
The renormalization is induced (to leading order) by low dimension
couplings symmetric under the extended continuous symmetry.
Thus the renormalization group masses and the potential will be symmetric
under the extended continuous symmetry.

However, this potential could be obtained in several other ways.
One particularly interesting possibility would be to generate the
potential
from strong coupling dynamics symmetric under the extended continuous
symmetry, allowing the inflaton to be intimately connected with the strong
coupling dynamics that presumably also generates the vacuum energy that
drives the inflation.
Normally it would be difficult to control the inflaton's mass in such a
context, but here it is protected by the discrete gauge symmetries.

In this paper we use the non-abelian discrete symmetry
$\Delta$(96) $\subset$ SU(3) described in the Appendix.
However, many other choices for the non-abelian discrete symmetry are
possible; for example, one could use non-abelian discrete subgroups of
SU(2) which would lead to more minimal models.
We use $\Delta$(96) simply for ease of model building.

To build a model one makes a suitable choice of gauge group and
representations.
The symmetries strongly constrain the allowed terms in the superpotential
and Kahler potential.
The resulting effective field theory is determined by the gauge
symmetries, the representations, the couplings, and the supersymmetry
breaking parameters.

The supersymmetry breaking parameters are in principle determined by
the supersymmetry breaking and how it is mediated to the relevant fields.
If supersymmetry is broken at a scale $F$, then one expects to induce
a vacuum energy $V_0 \sim |F|^2$, scalar mass squareds of order
$|F|^2/M^2$, where $M$ parametrizes the strength of the mediation and
should be somewhere in the range
$|F|^{1/2} \lesssim M \lesssim M_{\mathrm{Pl}}$, etc.
However, in our vacuum today supersymmetry is known to be broken at
a scale $|F|^2 \gsim \mathrm{TeV}^4$ but the vacuum energy is known to be
$\Lambda \sim 10^{-59} \, \mathrm{TeV}^4$.
This fine tuning of the cosmological constant will be transferred to
$V_0$ if vacuum supersymmetry breaking is the dominant supersymmetry
breaking in our effective field theory, allowing $V_0 \lesssim |F|^2$.
Note that $V_0$ itself breaks supersymmetry and so we can not have
$V_0 > |F|^2$.

However, as we do not wish to restrict ourselves to one particular model
of supersymmetry breaking, we do not specify $F$ or $M$ but rather just
consider the supersymmetry breaking parameters that appear in our low
energy effective theory and apply the appropriate constraints to them.
Our treatment is very similar to that of the Minimal Supersymmetric
Standard Model.
Specifically, we have no constraint on $V_0$ and require the scalar
mass squareds to be greater than the greater of $V_0 / M_{\mathrm{Pl}}^2$
and $ \mathrm{TeV}^4 / M_{\mathrm{Pl}}^2$.
The mechanism discussed in this paper is directed at protecting the mass
of the inflaton from this relatively large value.

Other parameters in our effective field theories include couplings,
which can be both dimensionless and dimensionful
(for example in the first model below $\lambda$ is dimensionless,
while $\sigma$ has dimensions of inverse mass). 
The dimensionless parameters should be of order one, unless protected
by some symmetry, but the dimensionful couplings go as some
inverse mass parameter in the theory, related to the fields that have
been integrated out and are thus not specified.
These dimensionful parameters can have a very wide range.
In the units we adopt, $\mpl \equiv 1$, a coupling $\sigma \sim 1/M$
for $M < \mpl$ will obey $\sigma > 1$ and can be very large.
However, the mass scales integrated out should be larger than the values
of the dynamical fields, \mbox{i.e.} the dynamical fields should have
values small enough for the effective field theory expansion to remain
valid, \mbox{i.e.} for given couplings, successively higher order terms
should get smaller and smaller.
Note that this requires the field values to be $\ll \mpl$.

With a given lagrangian in hand, we then impose the constraints coming
from inflation.
The effective field theory needs to provide a potential flat
enough for slow-roll inflation to occur (flatness),
a way for it to end (exit) and viable predictions for the primordial
flucutuation amplitude \cite{cobe} and tilt, including the absence of
any large spikes in the spectrum on observable wavelengths \cite{spike}.
Inflation ends soon after violation of slow-roll.
These constraints will determine allowed ranges for the parameters
in the models.

\section{A hybrid model}\label{hybrid}

We choose the gauge symmetries and fields shown in
Table~\ref{hybridtable}.
The non-abelian discrete symmetry $\Delta$(96) is described in the
Appendix.
The model is anomaly free.

\begin{table}[h]
\begin{center}
\begin{tabular}{|c|cccc|}
\hline
& $\Phi$ & $\Psi$ & $\Upsilon$ & $\Xi$ \\
\hline
$\Delta$(96) $\subset$ SU(3)
& $\bf{3}$ & $\bf{3}$ & $\bf{3}$ & $\bf{3}$ \\
${\bf Z}_3$
& $1$ & $-1$ & $1$ & $1$ \\
U(1)
& $0$ & $0$ & $1$ & $-1$ \\
\hline
\end{tabular}
\end{center}
\caption{\label{hybridtable}
Symmetries and fields in the hybrid model.
${\bf 3}$ represents a fundamental representation of both the discrete
gauge symmetry $\Delta$(96) and its global extension to SU(3).}
\end{table}
For this choice of symmetries and fields, the most general superpotential
is
\be
W = \lambda \Phi \wedge \Upsilon \wedge \Xi
+ \frac{\sigma}{2} \sum_{a=1}^3 \Phi_a^2 \Psi_a^2
+ \frac{\rho}{2} \sum_{a=1}^3 \Psi_a^2 \Upsilon_a \Xi_a
\ee
plus dimension 6 and higher terms.
Here, and throughout most of the rest of the paper, we have set $\mpl=1$
(not $\mpl=\infty$!).
Some other sector breaks supersymmetry, and in our low energy effective
field theory gives rise to the following general supersymmetry breaking
terms:
\bea
\Vsb & = & V_0 + \tilde{m}_\Phi^2 \lt| \Phi \rt|^2
- m_\Psi^2 \lt| \Psi \rt|^2
+ m_\Upsilon^2 \lt| \Upsilon \rt|^2
+ m_\Xi^2 \lt| \Xi \rt|^2
\nn \\ && \mbox{}
- \lt( \mu_\lambda \Phi \wedge \Upsilon \wedge \Xi
+ \mu_\sigma \sum_{a=1}^3 \Phi_a^2 \Psi_a^2
+ \mu_\rho \sum_{a=1}^3 \Psi_a^2 \Upsilon_a \Xi_a
+ \mbox{c.c.} \rt)
\eea
plus dimension 6 and higher terms.
Here $\tilde{m}_\Phi^2 \lt( \lt| \Phi \rt| \rt)$ is the SU(3) symmetric
renormalization group running mass squared of $\Phi$ induced by the SU(3)
symmetric coupling $\lambda \Phi \wedge \Upsilon \wedge \Xi$ in the
superpotential.
We assume that $\tilde{m}_\Phi^2 \lt( \lt| \Phi \rt| \rt) \lt| \Phi
\rt|^2$
has a minimum at $\lt| \Phi \rt| = \Phi_0$.
We also assume that $m_\Psi^2 > 0$, $m_\Upsilon^2 > 0$, and $m_\Xi^2 > 0$.
As mentioned earlier, generically the masses squared have magnitude
greater than or equal to $V_0 $ due to supergravity corrections.
Recall that $V_0$ is the vacuum energy at this time which is not
necessarily equal to the scale of supersymmetry breaking.

We consider the minimum in field space corresponding to the background
with $\Upsilon = \Xi = 0$.
The symmetries guarantee that this background is an extremum and one can
verify explicitly that it is stable if
$\lt| \Phi \rt| > \lt| \mu_\lambda / \lambda^2 \rt|$ or
$\lt| \lambda \rt|^2 \lt( m_\Upsilon^2 + m_\Xi^2 \rt)
> \lt| \mu_\lambda \rt|^2$.
We assume that $\Phi$ is located in the neighborhood of
$|\Phi|=\Phi_0$ and replace the term
$\tilde{m}_\Phi^2 \lt( \lt| \Phi \rt| \rt) \lt| \Phi \rt|^2$ by
$m_\Phi^2 \lt( \lt| \Phi \rt| - \Phi_0 \rt)^2$.
The leading terms are now
\be
W = \frac{\sigma}{2} \sum_{a=1}^3 \Phi_a^2 \Psi_a^2
\ee
and
\be
\Vsb = V_0
+ m_\Phi^2 \lt( \lt| \Phi \rt| - \Phi_0 \rt)^2
- m_\Psi^2 \lt| \Psi \rt|^2
- \lt( \mu_\sigma \sum_{a=1}^3 \Phi_a^2 \Psi_a^2 + \mbox{c.c.} \rt)
\ee
Note that the $D$-term is zero.
The term $m_\Phi^2 \lt( \lt| \Phi \rt| - \Phi_0 \rt)^2$ will constrain
$\Phi$
to lie on the sphere $\lt| \Phi \rt| = \Phi_0$.
The lowest order terms in the potential are then
\be\label{hp}
V = V_0 + \sum_{a=1}^3 \lt[
\lt| \sigma \rt|^2 \lt| \Phi_a \rt|^4 \lt| \Psi_a \rt|^2
+ \lt| \sigma \rt|^2 \lt| \Phi_a \rt|^2 \lt| \Psi_a \rt|^4
- \lt( \mu_\sigma \Phi_a^2 \Psi_a^2 + \mbox{c.c.} \rt)
- m_\Psi^2 \lt| \Psi_a \rt|^2
\rt]
\ee
with the constraint $\lt| \Phi \rt| = \Phi_0$.

This is a hybrid inflation \cite{hybrid} type potential.
When
\be
\label{phicdef}
\lt| \Phi_a \rt| > \Phi_{\rm c}
\equiv \sqrt{ \frac{\alpha m_\Psi}{\lt| \sigma \rt|} }
\,, \hspace{2em} a = 1, 2, 3
\ee
$\Psi$ is constrained to zero, leaving the potential
\be\label{V1}
V = V_0
\ee
with the constraint $\lt| \Phi \rt| = \Phi_0$.
When one of the $\lt| \Phi_a \rt|$ drops below $\Phi_{\rm c}$, the
potential
becomes unstable to $\lt| \Psi_a \rt| \rightarrow \infty$.
This may cause inflation to rapidly end, see Section~\ref{fast}, or there
could be more inflation as $\lt| \Psi_a \rt| \rightarrow \infty$, see
Section~\ref{slow}.
We have assumed
\be\label{conphi}
\Phi_0 > \sqrt{3}\, \Phi_{\rm c}
\ee
The constant $\alpha$ is given by
\be
\alpha = \sqrt{ 1 + \lt( \frac{\lt| \mu_\sigma \rt|}
 {\lt| \sigma \rt| m_\Psi} \rt)^2 }
+ \frac{\lt| \mu_\sigma \rt|}{\lt| \sigma \rt| m_\Psi}
\ee
We expect $\lt| \mu_\sigma \rt| \lesssim \lt| \sigma \rt| m_\Psi$ so that
$\alpha \sim 1$.

The potential, Eq.~(\ref{V1}), is flat with respect to the Nambu-Goldstone
bosons $\Phi_a / \lt| \Phi \rt|$.
The gauge symmetries have forbidden any terms which might
produce a large mass for the inflaton.
However, the higher dimension terms in the Kahler potential and
superpotential that are allowed by our gauge symmetries but which
violate the approximate global symmetry,
and that we have neglected up to now, will generate a gentle
slope. 
The relevant higher dimension invariants are $\Phi_1^2 \Phi_2^2 \Phi_3^2$,
$\sum_a \lt| \Phi_a \rt|^4$, and
$\sum_{a \neq b} \lt| \Phi_a \rt|^2 \lt| \Phi_b \rt|^2$, which generate
the
terms
\be
W = \ldots + \frac{1}{2} \nu \Phi_1^2 \Phi_2^2 \Phi_3^2
\ee
and
\be
\Vsb =  \ldots + m_1^2 \sum_a \lt| \Phi_a \rt|^4
+ m_2^2 \sum_{a \neq b} \lt| \Phi_a \rt|^2 \lt| \Phi_b \rt|^2
- \lt( \mu_\nu \Phi_1^2 \Phi_2^2 \Phi_3^2 + \mbox{c.c.} \rt)
\ee

Now for $\lt| \Phi \rt| = \Phi_0$ we have
\be
\sum_a \lt| \Phi_a \rt|^4 = \Phi_0^4
- 2 \sum_{a \neq b} \lt| \Phi_a \rt|^2 \lt| \Phi_b \rt|^2
\ee
and so
\bea
\label{potparts}
V & = & V_0 + m_1^2 \Phi_0^4
+ m_K^2 \sum_{a \neq b} \lt| \Phi_a \rt|^2 \lt| \Phi_b \rt|^2
- \lt( \mu_\nu \Phi_1^2 \Phi_2^2 \Phi_3^2 + \mbox{c.c.} \rt)
\nn \\ && \mbox{}
+ \lt| \nu \rt|^2
\lt| \Phi_1 \rt|^2 \lt| \Phi_2 \rt|^2 \lt| \Phi_3 \rt|^2
\sum_{a  \neq b} \lt| \Phi_a \rt|^2 \lt| \Phi_b \rt|^2
\eea
where $m_K^2 \equiv m_2^2 - 2 m_1^2$.

We assume the terms derived from the non-holomorphic invariants dominate
over the ones derived from the holomorphic invariant.
This can be ensured either by adding extra symmetry to the model, which
could set $\nu = 0$, or just by being in the appropriate region of
parameter
space ($m_K^2 \gg \mu_\nu \Phi_0^2$).
We also require $m_1^2 \Phi_0^4 \ll V_0$.
In order for the non-holomorphic term to drive the inflaton towards the
hybrid exit to inflation, we require $m_K^2 > 0$.
Then
\be
V = V_0 + m_K^2 \sum_{a \neq b} \lt| \Phi_a \rt|^2 \lt| \Phi_b \rt|^2
\ee
with the constraint $\lt| \Phi \rt| = \Phi_0$.
For simplicity, we assume\footnote{If instead
$\lt| \Phi_1 \rt|^2 \ll \lt| \Phi_2 \rt|^2 \sim \lt| \Phi_3 \rt|^2$,
the dynamics of $\Phi_1$ and $\Phi_2$ do not decouple.}
$\lt| \Phi_1 \rt|^2 , \lt| \Phi_2 \rt|^2 \ll \lt| \Phi_3 \rt|^2$.
Then
\be\label{mk}
V = V_0 + m_K^2 \Phi_0^2 \sum_{a=1}^2 \lt| \Phi_a \rt|^2
\ee

Quantum corrections will also generate a small slope
\bea
V_{\rm 1loop} & = &
\frac{1}{64\pi^2} \mbox{Str} {\cal M}^4 \ln \frac{{\cal
M}^2}{\Lambda^2}
\\ & = &
\frac{1}{64\pi^2} \sum_{a=1}^3 \lt\{
\lt[ \lt| \sigma \rt|^2 \lt| \Phi_a \rt|^4 - m_\Psi^2
 + 2 \lt| \mu_\sigma \rt| \lt| \Phi_a \rt|^2 \rt]^2
\ln \frac{\lt| \sigma \rt|^2 \lt| \Phi_a \rt|^4 - m_\Psi^2
 + 2 \lt| \mu_\sigma \rt| \lt| \Phi_a \rt|^2}{\Lambda^2}
\rt. \nn \\ && \hspace{41pt} \lt. \mbox{}
+ \lt[ \lt| \sigma \rt|^2 \lt| \Phi_a \rt|^4 - m_\Psi^2
 - 2 \lt| \mu_\sigma \rt| \lt| \Phi_a \rt|^2 \rt]^2
\ln \frac{\lt| \sigma \rt|^2 \lt| \Phi_a \rt|^4 - m_\Psi^2
 - 2 \lt| \mu_\sigma \rt| \lt| \Phi_a \rt|^2}{\Lambda^2}
\rt. \nn \\ && \hspace{41pt} \lt. \mbox{}
- 2 \lt[ \lt| \sigma \rt|^2 \lt| \Phi_a \rt|^4 \rt]^2
\ln \frac{\lt| \sigma \rt|^2 \lt| \Phi_a \rt|^4}{\Lambda^2}
\rt\}
\\ & = &
\frac{\lt| \sigma \rt|^4}{64\pi^2} \sum_{a=1}^3 \lt\{
\lt[ \lt( \lt| \Phi_a \rt|^2 - \alpha^{-2} \Phi_{\rm c}^2 \rt)
 \lt( \lt| \Phi_a \rt|^2 + \Phi_{\rm c}^2 \rt) \rt]^2
\ln \frac{\lt| \sigma \rt|^2
 \lt( \lt| \Phi_a \rt|^2 - \alpha^{-2} \Phi_{\rm c}^2 \rt)
 \lt( \lt| \Phi_a \rt|^2 + \Phi_{\rm c}^2 \rt)}{\Lambda^2}
\rt. \nn \\ && \hspace{41pt} \lt. \mbox{}
+ \lt[ \lt( \lt| \Phi_a \rt|^2 - \Phi_{\rm c}^2 \rt)
 \lt( \lt| \Phi_a \rt|^2 + \alpha^{-2} \Phi_{\rm c}^2 \rt) \rt]^2
\ln \frac{\lt| \sigma \rt|^2
 \lt( \lt| \Phi_a \rt|^2 - \Phi_{\rm c}^2 \rt)
 \lt( \lt| \Phi_a \rt|^2 + \alpha^{-2} \Phi_{\rm c}^2 \rt)}{\Lambda^2}
\rt. \nn \\ && \hspace{41pt} \lt. \mbox{}
- 2 \lt| \Phi_a \rt|^8
\ln \frac{\lt| \sigma \rt|^2 \lt| \Phi_a \rt|^4}{\Lambda^2}
\rt\}
\eea
For $\lt| \Phi \rt|^2 = \Phi_0^2 \gg \Phi_{\rm c}^2$ and
$\lt| \Phi_1 \rt|^2 , \lt| \Phi_2 \rt|^2 \ll \lt| \Phi_3 \rt|^2$, this
gives
\bea
V_{\rm 1loop} = \frac{\lt( 4 - \alpha^2 - \alpha^{-2} \rt)}{16 \pi^2}
\ln \lt( \frac{\lt| \sigma \rt|^2 \Phi_0^4}{\Lambda^2} \rt)
\lt| \sigma \rt|^2 m_\psi^2 \Phi_0^2
\sum_{a=1}^2 \lt| \Phi_a \rt|^2
\eea
This can be absorbed into Eq.~(\ref{mk}) if
\be\label{conm}
m_K^2 \gtrsim \lt| \sigma \rt|^2 m_\Psi^2
\ee

We assume $\lt| \Phi_1 \rt| \ll \lt| \Phi_2 \rt|$ so that $\lt| \Phi_1
\rt|$
controls the end of inflation and so is the relevant degree of freedom.
Defining $\phi = \sqrt{2} \lt| \Phi_1 \rt|$, $\psi = \sqrt{2} \lt| \Psi_1
\rt|$,
and $\phi_{\rm c} = \sqrt{2}\, \Phi_{\rm c}$, and reintroducing the hybrid
exit terms, Eq.~(\ref{hp}), (with phases relaxed and irrelevant terms
dropped), we get our effective model of inflation
\be
V = V_0 + \frac{1}{2} m_K^2 \Phi_0^2 \phi^2
+ \frac{1}{2} \lt( \frac{1}{4} \lt| \sigma \rt|^2 \phi^4
- \lt| \mu_\sigma \rt| \phi^2 - m_\Psi^2 \rt) \psi^2
\ee
There are two possibilities for when astronomically observable scales
could
leave the horizon during inflation; either at $\phi > \phi_{\rm c}$ or at
$\phi < \phi_{\rm c}$.
The former requires a quick hybrid exit in order to avoid possible
problems
with a spike in the density perturbation spectrum at $\phi = \phi_{\rm c}$
\cite{spike}.
The latter occurs in the opposite limit of a slow exit.

\subsection{Fast exit}\label{fast}

Here astronomically observable scales leave the horizon when
$\phi > \phi_{\rm c}$.
The slow roll conditions are satisfied if $m_K^2 \Phi_0^2 \ll V_0$.
The number of $e$-folds until $\phi = \phi_{\rm c}$ is
\be
N = \int_t^{t_{\rm c}} H \,dt
\simeq \int_{\phi_{\rm c}}^\phi \frac{V}{V'} \,d\phi
= \frac{V_0}{m_K^2 \Phi_0^2} \ln \frac{\phi}{\phi_{\rm c}}
\ee
The COBE normalization gives
\be
\frac{V^{3/2}}{V'} = \frac{V_0^{3/2}}{m_K^2 \Phi_0^2 \phi}
= \frac{V_0^{3/2}}{m_K^2 \Phi_0^2 \phi_{\rm c}}
\exp \lt( - \frac{m_K^2 \Phi_0^2 N}{V_0} \rt)
= 6 \times 10^{-4}
\ee
Substituting in for $\phi_c=\sqrt{2} \Phi_c$ and using
Eq.~(\ref{phicdef}),
this can be rewritten
\be
V_0^{1/4} = 10^{-3}
\lt( \frac{\alpha}{\lt| \sigma \rt|} \rt)^{1/2}
\lt( \frac{m_\Psi^2}{V_0} \rt)^{1/4}
\lt( \frac{m_K^2 \Phi_0^2}{V_0} \rt)
\exp \lt( \frac{m_K^2 \Phi_0^2 N}{V_0} \rt)
\ee
The spectral index is
\be
n = 1 + 2 \frac{V''}{V} = 1 + \frac{2 m_K^2 \Phi_0^2}{V_0}
\ee

A quick hybrid exit avoids problems at $\phi \sim \phi_c$, caused by
$\psi$'s fluctuations leading to too large a spike in the density
perturbation
spectrum, by making the time at which inflation ends effectively controlled
by $\phi$'s classical motion rather than by $\psi$'s stochastic
fluctuations.
The rough idea is that $\psi$'s effective mass squared goes from
$\gg H^2 \sim V_0$ to $\ll - H^2 \sim - V_0$ in a time-scale short compared
with the Hubble time so that the stochastic fluctuations in $\psi$, which
do
actually cause the end of inflation, do not lead to large fluctuations in
the
number of e-folds of expansion, and so do not lead to large density
perturbations.
In terms of parameters this means
\be\label{fec}
\lt. \frac{d M^2_\psi}{dN} \rt|_{\phi=\phi_{\rm c}}
= \lt. \frac{d M^2_\psi}{d\phi} \rt|_{\phi=\phi_{\rm c}}
\lt. \frac{d\phi}{dN} \rt|_{\phi=\phi_{\rm c}}
= \frac{2 \lt( \alpha^2 + 1 \rt) m_\Psi^2 m_K^2 \Phi_0^2}{V_0}
\gg V_0
\ee
where $M^2_\psi = \frac{1}{4} \lt| \sigma \rt|^2 \phi^4
- \lt| \mu_\sigma \rt| \phi^2 - m_\Psi^2$
is the effective mass of $\psi$.

This constraint, when combined with the others mentioned above,
severely restricts the parameter space.
However, pushing things to the limit, one can still come up with
interesting
numbers.
For example, taking $\Phi_0 =10^{-3.5} \lt| \sigma \rt|^{-1}$,
$m_\Psi =10^{-8} \lt| \sigma \rt|^{-1}$, and $m_K =10^{-8}$ gives
$V_0^{1/4} =10^{-5} \lt| \sigma \rt|^{-1/2}$ and $n = 1.002$.
Taking $\lt| \sigma \rt| = 10^8$ would then give
$m_\Psi =10^{-16} \simeq 200 \GeV$ and
$V_0^{1/4} =10^{-9} \simeq 2 \times 10^{9} \GeV$.

\subsection{Slow exit}\label{slow}

When $\phi \simeq \phi_{\rm c}$, $\psi$'s mass is partially canceled
\footnote{This is similar to the scenario of Ref.~\cite{Ross} in which
the expectation value of a Nambu-Goldstone boson is used to cancel off
the mass of the inflaton.
Our scenario has very different parameters, which leads to different
terms dominating the potential when observable scales leave the horizon
during inflation.}
allowing $\psi$ to slow-roll in addition to $\phi$.
Here astronomically observable scales leave the
horizon when $\phi < \phi_{\rm c}$.
Define
$\varphi = \phi_{\rm c} - \phi$.
Then
\be\label{slowpot}
V = V_0 - m_K^2 \Phi_0^2 \phi_{\rm c} \varphi
- \frac{\lt( \alpha^2 + 1 \rt) m_\psi^2}{\phi_{\rm c}} \varphi \psi^2
+ {\cal O} \lt( \frac{\varphi^2}{\phi_{\rm c}^2} \rt)
\ee
Note that when $\varphi$ becomes of order $\phi_{\rm c}$, $\psi$'s mass is
no longer suppressed and inflation ends rapidly, if it has not already
ended.
Thus $\varphi \ll \phi_{\rm c}$ will be a good approximation during
inflation.
The slow-roll equations of motion are
\be
\label{srphi}
\frac{d\varphi}{dN} = - \frac{m_K^2 \Phi_0^2 \phi_{\rm c}}{V_0}
- \frac{\lt( \alpha^2 + 1 \rt) m_\psi^2}{V_0 \phi_{\rm c}} \psi^2
\ee
\be
\label{srpsi}
\frac{d\psi}{dN} =
- 2 \frac{\lt( \alpha^2 + 1 \rt) m_\psi^2}{V_0 \phi_{\rm c}}
\varphi \psi
\ee
where
\be
N = \int_t^{t_{\rm e}} H \,dt
\ee
is the number of $e$-folds until the end of inflation.
Once
\be\label{psigg}
\psi^2 \gg
\frac{m_K^2 \Phi_0^2 \phi_{\rm c}^2}{\lt( \alpha^2 + 1 \rt) m_\psi^2}
\ee
one can solve this system of equations to give
\be\label{A}
\frac{1}{2} \psi^2 = \varphi^2 + A^2
\ee
where $A$ is a constant.
Substituting this into Eq.~(\ref{srphi}) and integrating gives
\be\label{N}
N = \frac{V_0 \phi_{\rm c}}{2 \lt( \alpha^2 + 1 \rt) A m_\psi^2}
\lt[ \tan^{-1} \frac{A}{\varphi} - \tan^{-1} \frac{A}{\varphi_{\rm e}}
\rt]
\ee
Therefore, once $\varphi$ and $\psi$ have rolled to values much greater
than
$A$, we have
\be
\varphi \sim \frac{1}{\sqrt{2}} \psi
\sim \frac{V_0 \phi_c}{2(\alpha^2 +1) m_\psi^2 N}
\ee
Therefore, in terms of $N$, the condition Eq.~(\ref{psigg}) translates to
\be
\label{slowcons}
2 \lt( \alpha^2 + 1 \rt) N^2 m_\psi^2 m_K^2 \Phi_0^2 \ll V_0^2
\ee
\mbox{i.e.} the limit opposite to that of Eq.~(\ref{fec}) of the previous
section.

Because both $\varphi$ and $\psi$ are slow-rolling, we need to use the
method of Ref.~\cite{Misao} to calculate the density
perturbations.\footnote{Note
that the dangerous spike in the density perturbations
produced at $\phi \sim \phi_{\rm c}$, \mbox{i.e.} $\varphi \sim 0$, is
inflated to unobservably large scales by the inflation that occurs at
$\varphi>0$. Our direct calculation shows that the density perturbations
are
acceptable on observable scales.}
The physics behind this method is very intuitive.
Stochastic fluctuations in the scalar fields lead to perturbations in the
number of $e$-folds of expansion.
Perturbations in the number of $e$-folds of expansion then induce
curvature
perturbations.
Finally, once these curvature perturbations re-enter the horizon after
inflation, they are naturally reinterpreted as density perturbations.
Now from Eq.~(\ref{N})
\be
N = \frac{V_0 \phi_{\rm c}}{2 \lt( \alpha^2 + 1 \rt) m_\psi^2 \varphi}
\lt[ 1 - \frac{A^2}{3 \varphi^2}
 + {\cal O} \lt( \frac{A^4}{\varphi^4} \rt)
 + {\cal O} \lt( \frac{\varphi}{\varphi_{\rm e}} \rt) \rt] \; .
\ee
To calculate the change in $N$ as $\varphi$ and $\psi$ are changed,
one also needs to use from Eq.~(\ref{A}) that
\be
\frac{\partial A}{\partial \varphi} = - \frac{\varphi}{A}
\ee
and
\be
\frac{\partial A}{\partial \psi} = \frac{\psi}{2A}
\ee
that is, one also needs to take into account fluctuations between
trajectories characterized by a given value of $A$, as well as
fluctuations
along a given trajectory.
Therefore, including this,
\be
\frac{\partial N}{\partial \varphi} =
- \frac{2 \lt( \alpha^2 + 1 \rt) m_\psi^2 N^2}{3 V_0 \phi_{\rm c}}
\ee
\be
\frac{\partial N}{\partial \psi} =
- \frac{2 \sqrt{2} \lt( \alpha^2 + 1 \rt) m_\psi^2 N^2}{3 V_0 \phi_{\rm
c}}
\ee
The COBE normalisation is \cite{Misao,cobe}
\be
\frac{H}{2\pi} \sqrt{ \lt( \frac{\partial N}{\partial \varphi} \rt)^2
+ \lt( \frac{\partial N}{\partial \psi} \rt)^2 } = 6 \times 10^{-5}
\ee
Therefore
\be
V_0^{1/2} \sqrt{ \lt( \frac{\partial N}{\partial \varphi} \rt)^2
+ \lt( \frac{\partial N}{\partial \psi} \rt)^2 }
= \frac{2 \lt( \alpha^2 + 1 \rt) m_\psi^2 N^2}{V_0^{1/2} \phi_{\rm c}}
= 6 \times 10^{-4}
\ee
and so
\be
\label{cobecon}
V_0^{1/4} = 10^{-7} \lt| \sigma \rt|^{-1/2}
\lt( \frac{2\sqrt{\alpha}}{\alpha^2 + 1} \rt)
\lt( \frac{45}{N} \rt)^2
\lt( \frac{V_0}{m_\psi^2} \rt)^{3/4}
\ee
The spectral index is
\be
\label{slowspec}
n = 1 - \frac{4}{N}
\ee
This is the same as one would get if one had a potential of the
form $V = V_0 - a \phi^3$, for example Ref.~\cite{Ross}.
However, the two models can in principle be distinguished by the
fact that our model does not satisfy the single component inflaton
consistency condition $n_T = - b T/S$.
Here $n_T$ is the spectral index of the gravitational waves,
$b$ is a constant that depends on conventions, and
$S$ and $T$ are the amplitudes of the scalar perturbations and
the gravitational waves, respectively \cite{consistency}.
Instead we have
\be
n_T = - 3 b \frac{T}{S}
\ee
In practice, though, this will be impossibly difficult to measure.

An interesting feature of this model is that it can easily produce
inflation at very low scales; for instance, to take an
extreme example, one can get
$V_0^{1/4} = 10^{-14} \simeq 20 {\, \rm TeV}$ with
$m_\psi \sim 10^{-24}$ and $\sigma \sim 10^2$.
This would, for example, be a low enough scale to replace thermal
inflation
\cite{thermal}.
It would also make embedding the model in the MSSM, or modest extensions
thereof, plausible.
However, the low scale of inflation means that less inflation is needed
and
so observable scales leave the horizon at relatively small values of $N$.
This, combined with the relatively large factor of $4$ in
Eq.~(\ref{slowspec}), results in a spectral index $n$ which is
too small to agree with observations.
One can get a more viable spectral index, \mbox{i.e.} $n$ closer to $1$,
by
raising the scale of inflation; for instance taking $V_0^{1/4} \sim
10^{-8}$.
Other parameters are then constrained by Eqs.~(\ref{conphi}),
(\ref{conm}),
(\ref{slowcons}) and~(\ref{cobecon}).

\section{A mutated hybrid model}\label{mutated}

To get a mutated hybrid inflation model, one can instead take
the symmetries and field content shown in Table~\ref{mutatedtable}.

\begin{table}[h]
\begin{center}
\begin{tabular}{|c|cccccc|}
\hline
& $\Phi$ & $\Psi$ & $\Upsilon$ & $\Xi$
& $\Omega$ & $\Gamma$ \\
\hline
$\Delta$(96) $\subset$ SU(3)
& $\bf{3}$ & $\bf{3}$ & $\bf{3}$ & $\bf{3}$ & $\bf{3}$ & $\bf{3}$ \\
${\bf Z}_9$
& $3$ & $-1$ & $3$ & $3$ & $-1$ & $-1$ \\
${\bf Z}_3$
& $0$ & $0$ & $1$ & $-1$ & $0$ & $0$ \\
${\bf Z}'_3$
& $0$ & $0$ & $0$ & $0$ & $1$ & $-1$ \\
\hline
\end{tabular}
\end{center}
\caption{\label{mutatedtable}
Symmetries and fields in the mutated hybrid model.
${\bf 3}$ represents a fundamental representation of both
the discrete gauge symmetry $\Delta$(96) and its global
extension to SU(3).}
\end{table}

The most general superpotential is
\be
W = \lambda \Phi \wedge \Upsilon \wedge \Xi
+ \frac{\sigma}{3} \sum_{a=1}^3 \Phi_a \Psi_a^3
+ \frac{\rho}{3} \sum_{a=1}^3 \Phi_a \Psi_a \Omega_a \Gamma_a
\ee
plus higher dimension terms, and the most general supersymmetry
breaking terms are
\bea
\Vsb & = & V_0 + \tilde{m}_\Phi^2 \lt| \Phi \rt|^2
- m_\Psi^2 \lt| \Psi \rt|^2
+ m_\Upsilon^2 \lt| \Upsilon \rt|^2
+ m_\Xi^2 \lt| \Xi \rt|^2
+ m_\Omega^2 \lt| \Omega \rt|^2
+ m_\Gamma^2 \lt| \Gamma \rt|^2
\nn \\ && \mbox{}
- \lt( \mu_\lambda \Phi \wedge \Upsilon \wedge \Xi
+ \mu_\sigma \sum_{a=1}^3 \Phi_a \Psi_a^3
+ \mu_\rho \sum_{a=1}^3 \Phi_a \Psi_a \Omega_a \Gamma_a
+ \mbox{c.c.} \rt)
\eea
plus higher dimension terms.
$\Phi$'s mass squared acquires a $\Phi$ dependence from the
renormalization group running induced by the coupling
$\lambda \Phi \wedge \Upsilon \wedge \Xi$ in the superpotential.
Since this coupling is SU(3) symmetric, the $\Phi$ dependence induced by
it
will also be SU(3) symmetric, \mbox{i.e.}
$\tilde{m}_\Phi^2 = \tilde{m}_\Phi^2 \lt( \lt| \Phi \rt| \rt)$.
We assume $\tilde{m}_\Phi^2 \lt( \lt| \Phi \rt| \rt) \lt| \Phi \rt|^2$
has a
minimum at $\lt| \Phi \rt| = \Phi_0$.
The higher dimension, SU(3) asymmetric couplings will induce a small SU(3)
asymmetric $\Phi$ dependence in the potential.
These small quantum corrections will be considered later.

The potential is minimized for $\Upsilon = \Xi = \Omega = \Gamma = 0$.
We assume that $\Phi$ is located in the neighborhood of $|\Phi|=\Phi_0$
and replace $\tilde{m}_\Phi^2 \lt( \lt| \Phi \rt| \rt) \lt| \Phi \rt|^2$
by
$ m_\Phi^2 \lt( \lt| \Phi \rt| - \Phi_0 \rt)^2 $.

In this background, the model simplifies to
\be
W = \frac{\sigma}{3} \sum_{a=1}^3 \Phi_a \Psi_a^3
\ee
and
\bea
V & = & V_0 + m_\Phi^2 \lt( \lt| \Phi \rt| - \Phi_0 \rt)^2
- m_\Psi^2 \lt| \Psi \rt|^2
- \lt( \mu_\sigma \sum_{a=1}^3 \Phi_a \Psi_a^3 + \mbox{c.c.} \rt)
\nn \\ && \mbox{}
+ \lt| \sigma \rt|^2 \sum_{a=1}^3 \lt| \Phi_a \rt|^2 \lt| \Psi_a
\rt|^4
+ \frac{1}{9} \lt| \sigma \rt|^2 \sum_{a=1}^3 \lt| \Psi_a \rt|^6
\eea
This is a mutated hybrid inflation \cite{mutated} type potential.
During inflation $\Phi$ constrains $\Psi$ to small but non-zero values
\be\label{mutatedtraj}
\lt| \Psi_a \rt| =
\frac{\beta m_\Psi}{\sqrt{2}\, \lt|\sigma\rt| \lt| \Phi_a \rt|}
\ee
where
\be
\beta = \sqrt{ 1 + \lt( \frac{3 \lt| \mu_\sigma \rt|}
 {2\sqrt{2}\, \lt| \sigma \rt| m_\Psi} \rt)^2 }
+ \frac{3 \lt|\mu_\sigma\rt|}{2\sqrt{2}\, \lt|\sigma\rt| m_\Psi}
\ee
and we have neglected the $\lt|\Psi_a\rt|^6$ term.
The effective potential for $\Phi$ is therefore
\be
V = V_0 + m_\Phi^2 \lt( \lt|\Phi\rt| - \Phi_0 \rt)^2
- \sum_{a=1}^3 \frac{\beta^2 \lt(\beta^2 + 2\rt) m_\Psi^4}
{12 \lt|\sigma\rt|^2 \lt|\Phi_a\rt|^2}
\ee

In the limit
$ \lt| \Phi_1 \rt|^2 \ll \lt| \Phi_2 \rt|^2 + \lt| \Phi_3 \rt|^2 $
this simplifies to
\be
V = V_0 - \frac{\beta^2 \lt(\beta^2 + 2\rt) m_\Psi^4}
{12 \lt|\sigma\rt|^2 \lt|\Phi_1\rt|^2}
\ee
which is a mutated hybrid inflation potential \cite{mutated}.
During inflation $|\Phi_1|$, or more precisely the field corresponding to
the
trajectory Eq.~(\ref{mutatedtraj}), rolls to smaller values and eventually
rolls fast enough to end inflation.

Mutated hybrid inflation has a spectral index \cite{mutated}
\be
n = 1 - \frac{3}{2N} \sim 0.97
\ee
and the COBE normalisation gives
\be
V_0^{1/4} = \frac{10^{-5}}{\sqrt{\lt|\sigma\rt|}}
\lt(\frac{50}{N}\rt)^{3/4} \frac{m_\Psi}{V_0^{1/2}}
\ee

\section{Discussion and Conclusions}\label{con}

We have discussed a mechanism to obtain potentials flat enough for
slow-roll inflation in the presence of supergravity corrections,
and given a hybrid and mutated hybrid example.
Our context has been that of a low energy effective field theory.  Discrete
gauge symmetries are used to guarantee that Planck scale effects do
not destroy the flatness of the potential, which is determined by
the choice of gauge symmetries, representations, and signs of the
supersymmetry breaking masses.  Constraints on the viable models
we considered were related to the mutated or hybrid exits.
The exit had to be approached via the slow roll potential and
additionally not generate fluctuations inconsistent with observation.
As this is a only a first attempt at building models implementing this
mechanism, we hope and believe
it is likely that more elegant versions are possible.

One attractive feature of this way of obtaining inflation is
that in principle, the inflationary scales for the hybrid models can be
very low.  In the specific case we looked at, the spectral index
becomes unviably small as the scale of inflation is lowered, but we do not
have any reason to expect this to be a generic limitation for
these sorts of models.  Inflation at very low scales has several
advantages.  For example, it might obviate the need for a round
of thermal inflation \cite{thermal}, as mentioned above,
to solve the moduli problem.   In addition, due to the low
energy scales involved, the model might
have a simple relation to phenomenological particle theory
models such as the minimal supersymmetric standard model.
One might also be able to make some correspondence with the discrete gauge
symmetries used here to obtain flatness and the discrete symmetries
in various parts of the standard model and its supersymmetric extensions,
for example those used for fermion masses, to suppress flavour changing
neutral currents, or in certain grand unified theories.

It should be stressed that this model is in the context of an
effective field theory.  As a result, certain properties of
the more complete theory cannot be deduced from the effective theory
alone, as they are more model dependent than the inflationary
mechanism and its exit described here.  These include
the details of (pre)heating and
the value of the cosmological constant today.

On a related note,
we have not discussed constraints from gravitino production in
the cases where these models have a higher inflationary scale.
This is primarily because, aside from the low reheating temperature
case mentioned above, a short era of low scale inflation
is needed to dilute the moduli, and will serve to dilute
the gravitinos as well.  In addition, the amount of gravitino
production is strongly model dependent, and thus our effective field theory
does not necessarily contain enough information to predict it.
Future directions include implementing this idea for different
gauge groups, and embedding an effective theory with this
mechanism into a more complete model.

\section*{Appendix}
\label{Appendix}

$\Delta$(96) is the discrete subgroup of SU(3) with elements
\cite{symmetry}
\be
X_{mn} \equiv A_{mn} X_{00}
\ee
where
\be
A_{mn} \equiv
\lt( \begin{array}{ccc} i^m & 0   & 0        \\
                        0   & i^n & 0        \\
                        0   & 0   & i^{-m-n} \end{array} \rt)
\ee
and
\bea
X_{00} & \in & \lt\{
\lt( \begin{array}{ccc} 1 & 0 & 0 \\
                        0 & 1 & 0 \\
                        0 & 0 & 1 \end{array} \rt),
\lt( \begin{array}{ccc} 0 & 1 & 0 \\
                        0 & 0 & 1 \\
                        1 & 0 & 0 \end{array} \rt),
\lt( \begin{array}{ccc} 0 & 0 & 1 \\
                        1 & 0 & 0 \\
                        0 & 1 & 0 \end{array} \rt),
\rt. \nn \\ && \lt. \mbox{}
\lt( \begin{array}{ccc} 0  & 1 & 0 \\
                        -1 & 0 & 0 \\
                        0  & 0 & 1 \end{array} \rt),
\lt( \begin{array}{ccc} 1 & 0  & 0 \\
                        0 & 0  & 1 \\
                        0 & -1 & 0 \end{array} \rt),
\lt( \begin{array}{ccc} 0 & 0 & -1 \\
                        0 & 1 & 0  \\
                        1 & 0 & 0  \end{array} \rt)
\rt\}
\eea
It can be generated by
\be
\lt\{
\lt( \begin{array}{ccc} 0 & 1 & 0 \\
                        0 & 0 & 1 \\
                        1 & 0 & 0 \end{array} \rt),
\lt( \begin{array}{ccc} 0 & i & 0 \\
                        i & 0 & 0 \\
                        0 & 0 & 1 \end{array} \rt)
\rt\}
\ee

Let $\Phi_a$, $\Psi_a$, $\Upsilon_a$, $\Xi_a$, $\Omega_a$, and
$\Gamma_a$
transform as fundamental representations of $\Delta$(96),
where $a = 1, 2, 3$ labels the components of the representation.

The holomorphic invariants of $\Delta$(96) are
\be\label{inv3}
\Phi \wedge \Psi \wedge \Upsilon \equiv
\sum_{a,b,c} \epsilon_{abc} \Phi_a \Psi_b \Upsilon_c
\ee
\be
\sum_a \Phi_a \Psi_a \Upsilon_a \Xi_a
\ee
\be
\sum_{a \neq b \neq c \neq a}
\Phi_a \Psi_a \Upsilon_b \Xi_b \Omega_c \Gamma_c
\ee
plus dimension 7 and higher invariants.

Non-holomorphic invariants are
\be\label{inv2}
\Phi^\dagger \Psi \equiv \sum_a \Phi_a^* \Psi_a
\ee
\be
\sum_a \Phi_a^* \Psi_a^* \Upsilon_a \Xi_a
\ee
\be
\sum_{a \neq b} \Phi_a^* \Psi_b^* \Upsilon_a \Xi_b
\ee
plus dimension 5 and higher invariants.

Note that the lowest dimension holomorphic and non-holomorphic
invariants,
Eqs.~(\ref{inv3}) and~(\ref{inv2}),
are symmetric under the full continuous SU(3) group.

\subsection*{Acknowledgements}

This work was supported by the DOE and the NASA grant NAG 5-7092 at
Fermilab and by Grant No. 1999-2-111-002-5 from the
interdisciplinary Research Program of the KOSEF.
JDC was supported in part by NSF-PHY-9800978 and NSF-PHY-9896019.
EDS thanks M. White and R. Leigh at
UIUC for hospitality while this work was begun, JDC thanks
the Aspen Center for Physics for hospitality while this work was in progress,
and we both thank the Santa Fe 99 Workshop on Structure Formation
and Dark Matter for hospitality while this work was mostly completed.
JDC is grateful to Martin White for discussions and Jochen
Weller for a helpful question, and we both thank M. Dine for helpful
suggestions on the draft.

\end{document}